*Review*

# A Forward, Analytic, Differentiable, Geometric (but Inflexible) Lens Model

Paul L. Schechter

MIT Kavli Institute, Massachusets Institute of Technology, 70 Vassar Street, Cambridge, MA 02139, USA; schech@mit.edu

**Abstract**

We anticipate that hundreds of thousands of distant, strongly gravitationally lensed sources will be detectable with European Space Agency's (ESA) Euclid mission and the Rubin Observatory Legacy Survey of Space and Time. We consider the virtues and shortcomings of the Singular Isothermal Elliptical Potential ($SIEP$) with Parallel External Shear ($XS_{\parallel}$) for these systems. Its principal virtue is that it admits an analytic forward model that gives image positions and magnifications as functions of the source position (and shape for extended sources). Preliminary experiments suggest a speed-up of a factor in excess of 10,000 compared with conventional models that instead map from the image plane to the source plane and require iteration to converge upon a unique source. A second virtue is that the Witt–Wynne geometric representation of $SIEP + XS_{\parallel}$ permits the quick visual verification of the model's adequacy for a particular lensed system. Unfortunately, the model's strictly elliptical lens equipotential is inconsistent with strictly elliptical surface mass density contours. The Witt–Wynne construction might nonetheless yield a sufficiently good first approximation to accelerate convergence to one's preferred lens model.

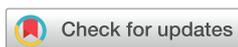

## 1. Introduction

The Singular Isothermal Elliptical Potential (SIEP) with Parallel External Shear ($XS_{\parallel}$) has four features that offer substantial advantages in searching for and modeling the hundreds of thousands of distant gravitationally lensed sources forecast for ESA's Euclid mission and for the Rubin Observatory Legacy Survey of Space and Time (LSST).

We begin by introducing the model's virtues. In Section 2, we give a thorough description of $SIEP + XS_{\parallel}$. In Section 3, we illustrate the model's shortcomings. In Section 4, we examine the suitability of $SIEP + XS_{\parallel}$ for use in searching for gravitationally lensed systems. In Section 5, we suggest how the model can be used as a first approximation to accelerate modeling with more sophisticated alternatives.

*1.1. Why Forward?*

Roughly 200 papers posted on astro-ph during 2025 use forward modeling (as judged from their abstracts), either implying or explicitly asserting that forward modeling is preferable to inverse modeling. Among the many astronomical phenomena and technical challenges addressed are gamma ray bursts, stellar oscillations, stellar metallicity distributions, spectral line fitting, and point spread function fitting.

Forward modeling applies inherently approximate theories to pristine observations, leaving the data untouched except, in some cases, to remove instrumental signatures.

Inverse modeling is eschewed because unmodeled phenomena bias the resulting best model in unknown ways.

In the present context of strong gravitational lensing, forward modeling begins with a parameterized model for the gravitational potential of the lens (position in sky, depth of potential, shape) and the position and unlensed flux of the source (and perhaps its shape); the positions, lensed fluxes (and perhaps shapes) of multiple images of the source are derived.

*1.2. Why Analytic?*

Under most circumstances, an analytic expression for an observable quantity drawn from a model can be evaluated more rapidly than using numerical techniques, with the possible exception of interpolating over a tabulated result. Predictions from analytic expressions may also facilitate debugging.

*1.3. Why Differentiable?*

Differentiability permits analytic (as opposed to numerical) differences to be used in gradient and Newton–Raphson searches, accelerating convergence when optimizing a fit.

*1.4. Why Geometric?*

Models that admit a geometric solution offer insights into the underlying physics that complement algebraic and algorithmic solutions and verbal descriptions. The combination of Kepler's ellipses and equal areas law may be the best-known example of this complementarity.

*1.5. What Imperfections?*

The above-mentioned benefits come at a cost. Our forward, analytic, differentiable geometric lens model assumes a singular isothermal elliptical *potential* with parallel shear. It is limited in describing elliptical surface mass density profiles, power laws other than isothermal, non-singular isothermals, and systems in which the shear is not parallel to the elliptical lens. Most lens modelers adopt one or more of these precluded features for good reasons.

The word model first came into use to describe scaled-down representations of objects that preserved large-scale features. In what follows, we illustrate our model's shortcomings. Our summary in Section 6 takes the form of a question addressed to the reader. Do our model's virtues compensate for its imperfections?

## 2. Singular Isothermal Elliptical Potential with Parallel Shear $SIEP + XS_{\parallel}$

The virtues we claim for the $SIEP + XS_{\parallel}$ model originate from our assumption of an elliptical potential rather than an elliptical surface mass density. Witt [1] discovered that when concentric similar elliptical equipotentials produce four images, one image of each parity lies on each branch of a rectangular hyperbola, one having orthogonal asymptotes. The equipotentials are aligned with these asymptotes, and their common center lies on what we call the "primary branch." Witt also found that the four images produced by a singular isothermal spherical potential with external shear likewise lie on the branches of a rectangular hyperbola, as does the center of the isothermal.

Wynne [2] discovered that for a singular *elliptical* isothermal potential, the four images lie on an ellipse, with the source at its center, whose axis ratio is the same as that of the potential but whose major axis is perpendicular to the potential's long axis. The source *also* lies on the primary branch of Witt's hyperbola.

Luhtaru [3] showed that adding shear parallel to the equipotentials gave a one-dimensional family of potentials for which images form where Wynne's ellipse intersects Witt's hyperbola.

The deflection of a ray from a QSO by a thin lens by $\vec{r} - \vec{r}_{QSO}$ (where $\vec{r}$ is the angular position in the sky) is found from the lens equation [4] to be the gradient of the dimensionless potential $\Phi_{2D}$. For $SIEP + XS_{\parallel}$, we have

$$\underbrace{\vec{r} - \vec{r}_{QSO}}_{\text{deflection}} = \vec{\nabla} \left\{ \underbrace{b \left[ (x - x_{\text{lens}})^2 + \frac{(y - y_{\text{lens}})^2}{q^2} \right]^{\frac{1}{2}}}_{\text{singular isothermal elliptical potential}} - \underbrace{\frac{\gamma}{2} \left[ (x - x_{QSO})^2 - (y - y_{QSO})^2 \right]}_{\text{external shear}} \right\}, \quad (1)$$

where $b$ is the semi-major axis of Wynne's ellipse, $q$ is its axis ratio, and $x_{lens}$ and $y_{lens}$ are the components of the angular position $\vec{r}_{lens}$. For the sake of simplicity, the $x$ axis is taken to be parallel to the long axis of the potential. The contribution of the shear $\gamma$ to the potential has no cross term and is therefore parallel to the elliptical isothermal.

The deflection has a modulus and a direction. The direction yields Witt's hyperbola for $SIEP + XS_{\parallel}$,

$$\frac{y - y_{QSO}}{x - x_{QSO}} = \underbrace{\left(\frac{1+\gamma}{1-\gamma}\right) \frac{1}{q^2}}_{1/Q_{\text{hyp}}^2} \left( \frac{y - y_{\text{lens}}}{x - x_{\text{lens}}} \right) \quad (2)$$

with an effective axis ratio $Q_{hyp} \equiv q\sqrt{(1-\gamma)/(1+\gamma)}$.

The modulus of the lens equation yields the corresponding Wynne ellipse,

$$(x - x_{QSO})^2 + \underbrace{\left(\frac{1-\gamma}{1+\gamma}\right)^2 q^2}_{1/Q_{\text{ell}}^2} (y - y_{QSO})^2 = b^2 \left(\frac{1}{1+\gamma}\right)^2 \Leftarrow \text{Wynne's rmellipse} \quad (3)$$

with an effective axis ratio $Q_{ell} \equiv (1+\gamma)/(1-\gamma)\gamma$. These results are drawn from [3].

## 3. Shortcomings of $SIEP + XS_{\parallel}$ Model

*3.1. Residuals*

Figure 1 shows the application of the Witt–Wynne construction to a sample of twelve quadruply lensed QSOs selected for observation during HST cycle 25. The position of the lensing galaxy was not used in determining either Witt's hyperbola or Wynne's ellipse.

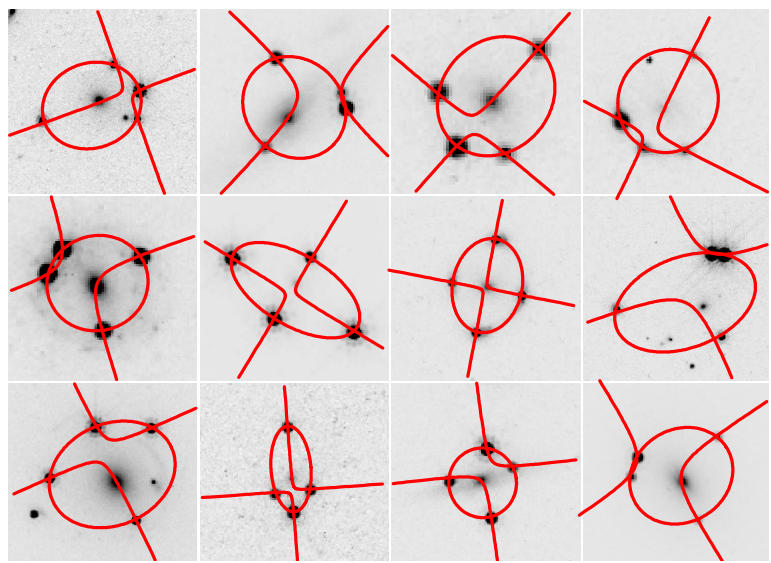

**Figure 1.** The red curves show Witt hyperbolae and Wynne ellipses overlaid on HST images of quadruply lensed quasars. The QSO images lie close to the points where the ellipses cross the hyperbolae.

The model has 7 parameters and 8 constraints (the coordinates of the four images). Rather than apportioning the residuals between the ellipse and the two branches of the hyperbola, the ellipses were chosen to pass precisely through the images, as was the "secondary" branch of Witt's hyperbola. The two images associated with the primary branch are very slightly displaced from it. Our fitting scheme, described in Appendix A, permits the complete calculation of a model with only 3 matrix inversions, the largest of which is $3 \times 3$.

The residuals are largest for the cluster J0818-26 shown in the rightmost panel of the middle row. Several member galaxies lie close to Wynne's ellipse. Random quartets of points produce substantially larger residuals.

Though the position of the central galaxy was not used in the construction, the primary branch of Witt's hyperbola passes through it in every case except for the cluster. The position of the lensing galaxy along the primary branch depends upon the relative contributions of shear and the ellipticity of the isothermal to the quadrupole moment of the potential.

*3.2. Non-Elliptical Mass Profiles*

A two dimensional version of the divergence theorem applies to the dimensionless lens potential. The Laplacian of the potential with respect to the angular position gives twice the convergence, $\kappa$, of the lens' surface mass density [4]. Application to the singular isothermal elliptical potential does *not* give an elliptical surface mass density. Luhtaru et al. [3] rediscovered an expression (see also [5]) for the axis ratio of the SIEP that most closely reproduces the image configuration from a singular elliptical mass and found

$$q_{pot} = \frac{\tan^{-1}\left(\sqrt{1-q_{mass}^2}/q_{mass}\right)}{\tanh^{-1}\left(\sqrt{1-q_{mass}^2}\right)} \sim q_{mass}^{1/3} \quad (4)$$

The "elliptical" galaxies of Hubble's classification scheme [6] have surface brightness profiles that are only approximately elliptical, with fourth- and higher-order multipoles that deviate from strict ellipticity. Rather than introduce additional free parameters to quantify additional multipoles in the surface mass density, many modelers take elliptical equidensity contours to be the least objectionable option.

*3.3. Non-Parallel Shear*

Holder and Schechter [7] carried out simulations from which they interpreted an apparent excess of quadruply lensed systems in the CLASS sample as the consequence of stronger shear than had previously been assumed. Luhtaru et al. [3] concluded that for the case of a quadruply lensed quasar with only a single bright galaxy inside Wynne's ellipse, shear typically contributes twice as much to its flattening as the ellipticity of the lens' halo.

The non-ellipticity of the lensing halos, their non-parallelism, and measurement errors combine to give the residuals in Figure 1.

*3.4. (In-)Appropriateness for Modeling Clusters of Galaxies*

As judged from weak lensing, many clusters of galaxies have irregular mass distributions that are considered the products of recent or ongoing mergers. Hanna et al. [8] succeeded in using Witt–Wynne with quadruply lensed background galaxies to determine the center of the potential of the very rich cluster Abell 1689 with 2 arcsecond accuracy, but, more typically, the $SIEP + XS_\parallel$ model is less well-suited to cluster lenses, as seen in the rightmost panel of the middle row of Figure 1.

## 4. Suitability of Witt–Wynne Construction for Searches

*4.1. Point Sources*

Searches for quadruple lens systems can be carried out on either catalog entries or pixellated cutouts of the observed sky. (While the Witt–Wynne construction was developed to model quads, it also produces doubles when the secondary branch of Witt's hyperbola does not cross Wynne's ellipse. Pairs of catalog entries can be supplemented with the position of the lensing galaxy (taken to be the center of the potential), the fluxes of the point images, the position angle of the lensing galaxy, and the assumption of achromaticity.) The blind search of the Gaia catalog by Delchambre et al. [9] may be the best example of the former. They carried out a random forest search on 80,000 quartets of Gaia positions calibrated using $10^8$ simulated systems. The Witt–Wynne position-fitting scheme described in Section 3.1 above modeled those quartets in less than 10 s on the author's 2012 laptop. While such systems are subject to chromatic micro-lensing, a mild prior on achromaticity consistent with Weisenbach's [10] "worst case" scenarios helps remove false positives.

A major shortcoming of catalog searches is that each of the four images must, by itself, satisfy the criteria for inclusion in the catalog. Suppose a $5\sigma$ detection is required. All four of the lensed images would have to satisfy that criterion. A system with four $4\sigma$ detections would not. By contrast, such a system would produce a high significance detection if the cutout pixels were directly modeled as a quadruple system.

The Witt–Wynne construction is unique among the lens models permitting forward, analytic, differentiable modeling of cutout pixels, and relatively straightforward if the source produces point images. Falor and Schechter [11] present a closed-form scheme for modeling the latter, at whose center lies a single equation that gives the angular position on Wynne's ellipse for all four images.

Let $\rho$ and $\theta$ be the polar coordinates of the four images for a nearly circular, $q_{pot} \approx 1$, singular iosthermal elliptical potential. Falor solve a quartic equation and find four images at

$$\begin{aligned}2e^{i\theta} &= W \hat{\pm} \sqrt{u + W^2} \\ &\tilde{\pm} \sqrt{(W \hat{\pm} \sqrt{u + W^2})^2 - 2\left(u \hat{\pm} \sqrt{u^2 + 4}\right)}\end{aligned} \quad (5)$$

where the symbols $\hat{\pm}$ and $\tilde{\pm}$ reflect two distinct invocations of the quadratic formula. Here, $W \equiv x_{hyp} + iy_{hyp}$ is the center of Witt's hyperbola in a coordinate system centered on Wynne's ellipse, and $\overline{W}$ is its complex conjugate. The complex quantity $u$ is the solution of a cubic equation

$$u^3 + 4(1 - W\overline{W})u + 4(W^2 - \overline{W}^2) = 0. \quad (6)$$

Wrapped around this is a linear coordinate transformation into a system that renders Wynne's ellipse nearly circular and a transformation back into the original coordinate system. An implementation of this scheme by Falor was found by Ludovic Delchambre (private communication), who found the code to be 20,000 times faster than the code he had fine-tunedt to generate SIEP models with arbitrary shear as a training set for Gaia. It was 450,000 times faster than `gravlens`, a widely used general purpose code written by Keeton [12]. While the speed-up achieved with Falor's algorithm varies from one code to the next, these two benchmarks give a sense of what may be possible.

The forward nature of Falor's solution permits the acceleration of the follow-up of lensed supernovae such as SN 2025wny [13,14], for which the host was previously known to be quadruply imaged. A pre-computed Witt–Wynne model for the host projects a diamond caustic onto the source plane. Projecting a transient backward onto that caustic, one sees whether it produces four or only two images (as it might if the transient is offset from the host). The forward application of Falor's method then immediately predicts (or post-dicts) the positions of trailing and leading images, their time delays, and their magnifications.

In the case of SN 2025wny, one might have realized that the triggering event was the trailing image and that the corresponding brightening of the leading image occurred when the object was behind the Sun (Schechter and Lu, in preparation). Perhaps one would have chosen to wait rather than trigger the associated target of opportunity programs.

*4.2. Extended Sources*

Models for extended sources involve at least three more parameters than those for point sources. These might be a semi-major axis, a position angle, and an axis ratio. In our forward scheme, the source must be sampled at a finite number of points. One starts with a preliminary model, predicts the observed flux for each observed pixel, and then minimizes a weighted sum of the residuals in the image plane. If some pixel in the image plane is irreparably corrupted (e.g., by a diffraction spike), one gives that pixel zero weight.

One might alternatively assume a model for the source and map each pixel backward to the source plane, minimizing a weighted sum of residuals in that plane. In the absence of a perfect lens model, noise at some position relative to the centers of the four images (e.g., diffraction spikes) is mapped to different pixels in the source plane. Correcting for such imperfections is less straightforward than in the forward case.

*4.3. Searching by Convolution*

Four decades have passed since the first use of convolution with matched filters to identify astronomical objects [15]. The number of available pixels on a typical CCD has grown by a factor of 100. These have in turn been combined in mosaics with as many as 200 CCDs. Those first convolutional searches were carried out in regions of high stellar density. The same technique can now be used to search for extended sources across the whole sky.

Unlike stars, extended sources come in a wide variety of shapes, requiring a large library of matched filters. These are often generated synthetically. The main strength of convolutional matching is its speed. The shortcoming is the computation required to generate the filters and training sets.

Metcalf et al. [16] described a two-part "Strong Gravitational Lens Finding Challenge", the first of which simulated ground-based observations in four filters and the second of which simulated satellite observations in a single filter. In both cases 100,000 lines of sight were sampled, of which roughly 50% produced detectable lensing. Of the 24 entries in the challenge, 17 involved convolutional techniques. Had Falor's algorithm been available, the designers of the challenge might have accelerated the synthesis of their lines of sight, as suggested by the speed-up reported by Delchambre.

## 5. Suitability of Witt–Wynne Construction for Initial Modeling

At a 2025 meeting in Liege on strong lensing, a prominent lens modeler outlined the steps in their particular approach, which involved first adopting an initial model and second developing a more complex model that addressed the particular science case being addressed. They went on to say "once you have an initial model, it is easy to fit a complex lens model".

The better the initial guess in a Markov Chain Monte Carlo optimization, the more rapidly it will converge, speeding up the computation of the more complex model. The Witt–Wynne construction might suffice for an initial model.

Modelers must ask themselves whether the gain in wall clock and CPU time resulting from initial $SIEP + XS_\parallel$ modeling warrants the up-front cost of embedding Witt–Wynne in their code.

## 6. Do the Witt–Wynne Strengths Compensate for Its Imperfections?

In the preceding sections, we described both the greatest virtues and the most serious shortcomings of the Witt–Wynne model. We ask the reader to take 60 s to look back at our section headings and at Figure 1 and to record (on a screen or a piece of paper) either the word "yes" or the word "no" to the question asked by this section.

**Funding:** This research was supported by the author's personal resources.

**Data Availability Statement:** The observations used in Figure 1 were carried out with the NASA/ESA Hubble Space Telescope and obtained from the Space Telescope Science Institute, which is operated by the Association of Universities for Research in Astronomy, Inc. under NASA contract NAS 5-26555. They were carried out under program HST-GO-15652.

**Acknowledgments:** The author thanks Ludovic Delchambre for working with C. Falor to test the speed of his algorithm.

**Conflicts of Interest:** The author declares no conflicts of interest.

## Appendix A. Fast Inverse Solution for SIEP and SIS + XS Models Using Positions of Four Images

Given our preference for forward modeling, we emphasized the gains that might be achievable using $SIEP + XS_\parallel$ in this mode. But it also permits fast inverse modeling from quadruply lensed image positions back to the source plane, especially if one is willing to constrain two of the image positions to lie on the "secondary branch" of Witt's hyperbola.

*Appendix A.1. A Geometric Model from Image Positions*

The *x* and *y* positions of the four images give eight constraints. Three of the seven model parameters are, for illustrative purposes, attributed to Wynne's ellipse: its orientation, and its semi-major-axis and its axis ratio. Three are attributed to Witt's hyperbola: the coordinates of its center and its semi-major axis. The seventh parameter the gives the position of the center of Wynne's ellipse along the "primary" branch of Witt's hyperbola, to which it is constrained. Wynne's ellipse is also constrained to have symmetry axes parallel to the asymptotes of Witt's hyperbola.

The process for obtaining the Witt–Wynne model proceeds as follows:

1. Find the "rectangular" hyperbola passing through the four image positions by inverting a $3 \times 3$ matrix.
2. Use the orientation of Witt's hyperbola to find the image coordinates in a frame whose axes are parallel to the asymptotes of the hyperbola.
3. Find the ellipse passing through the four image positions whose semi-major and semi-minor axes are parallel to the asymptotes of Witt's hyperbola by inverting a $3 \times 3$ matrix.
4. Identify the two images subtending the smallest angle from the center of the amplitude ellipse.
5. Find the "rectangular" hyperbola parallel that of the first step that passes through these two images and the center of Wynne's ellipse by inverting a $2 \times 2$ matrix.

Note that the dimensions quoted for the matrix inversions above are all smaller (e.g., $3 \times 3$ instead of $4 \times 4$) than one might first think. This is accomplished by taking one of the fitted points as the temporary origin of the coordinate system.

Constraining the two closest images to lie on the secondary branch of Witt's hyperbola guarantees a model that produces four and not just two images. The other two images need not lie at the intersection of the hyperbola and the ellipse. The worst case in Figure 1 is for the cluster lens in the rightmost center row.

*Appendix A.2. Connecting Geometric Construction to Physics*

Witt [1] demonstrates that his hyperbola works *separately* for both the case of the $SIEP$ and for that of the singular isothermal sphere with external shear ($SIS + XS_{\parallel}$).

The scheme described in the previous subsection is purely geometric: it refers neither to the source nor to the lens potential. But, working forward from the deflection Equation (1), one finds that, for the case of zero shear, the major axis of Wynne's ellipse is perpendicular to the major axis of the elliptical potential. Its center lies on Witt's hyperbola. The latter is also true for the case of zero ellipticity and finite shear, but the minor axis of Wynne's ellipse now points in the direction of the perturbing mass (if there is only one). The shear is approximately equal to the semi-ellipticity of Wynne's ellipse.

The Witt–Wynne geometric model does not require either the shear or the ellipticity of the lens to be zero. Luhtaru et al. [3] showed that there is a one-dimensional family of models when the shear is parallel to the ellipticity of the lens. The position of the center of the potential determines which component dominates the shape of Wynne's ellipse.